\documentclass[12pt, a4paper]{article}

\usepackage[english]{babel}

\usepackage[letterpaper,top=2cm,bottom=2cm,left=3cm,right=3cm,marginparwidth=1.75cm]{geometry}

\usepackage{amsmath}
\usepackage{authblk}
\usepackage{graphicx}
\usepackage{subcaption}
\usepackage[colorlinks=true, allcolors=blue]{hyperref}
\usepackage{bm}

\usepackage[numbers,sort&compress]{natbib}

\graphicspath{{figures/}}

\title{Gravitational Waves from Preheating in Inflation with Weyl Symmetry}
\author[1]{Wei-Yu Hu\footnote{weiyuhu@stu.pku.edu.cn}}
\author[3]{Qing-Yang Wang\footnote{wangqingyang19@mails.ucas.ac.cn}}
\author[1,2]{Yan-Qing Ma\footnote{yqma@pku.edu.cn}}
\author[3,4,5]{Yong Tang\footnote{tangy@ucas.ac.cn}}
\affil[1]{\small School of Physics, Peking University, Beijing 100871, China}
\affil[2]{\small Center for High Energy Physics, Peking University, Beijing 100871, China}
\affil[3]{\small University of Chinese Academy of Sciences (UCAS), Beijing 100049, China}
\affil[4]{\small School of Fundamental Physics and Mathematical Sciences, Hangzhou Institute for Advanced Study, UCAS, Hangzhou 310024, China}
\affil[5]{\small International Centre for Theoretical Physics Asia-Pacific, UCAS, Beijing 100190, China}

\date{\today}
\begin{document}
\maketitle

\begin{abstract}
Inflation with Weyl scaling symmetry provides a viable scenario that can generate both the nearly scaling invariant primordial density fluctuation and a dark matter candidate. Here we point out that, in additional to the primordial gravitational waves (GWs) from quantum fluctuations, the production of high-frequency GWs from preheating in such inflation models can provide an another probe of the inflationary dynamics. We conduct both linear analytical analysis and nonlinear numerical lattice simulation in a typical model. We find that significant stochastic GWs can be produced and the frequency band is located around $10^8$ Hz $\sim$ $10^9$ Hz, which might be probed by future resonance-cavity experiments.
\end{abstract}

\section{Introduction}

Cosmic inflation describes a quasi-exponential expansion stage in the early Universe and was conceived as a solution to the shortcomings of the hot Big Bang theory,
such as the cosmological horizon problem and flatness problem \cite{Starobinsky:1980te,Guth:1980zm,Linde:1981mu,Albrecht:1982wi}. It also provides a physical mechanism to generate the nearly scaling-invariant primordial perturbation, which is the seed of large scale structure (LSS) \cite{Mukhanov:1981xt}. However, the exact mechanism of inflation is still unclear, though various models of inflation have been constructed~\cite{Lyth:1998xn}, awaiting for future experimental tests. 

Recently, several scenarios based on the approximate Weyl scaling symmetry during inflation were proposed to provide viable inflation and a dark matter candidate~\cite{Tang:2018mhn, Tang:2019olx, Tang:2019uex, Tang:2020ovf, Tang:2021lcn, Wang:2022ojc, Wang:2023hsb}, both of which are essential for the LSS. Many studies motivated by Weyl symmetry have been explored in gauge theory of quantum gravity~\cite{Wu:2015wwa, Wu:2017urh}, induced gravity~\cite{Zee:1978wi, Adler:1982ri, Fujii:1982ms, Salvio:2014soa, Oda:2020yyv, Karananas:2021gco}, extensions of the standard model of particle physics~\cite{Cheng:1988zx, Hur:2011sv, Holthausen:2013ota, Foot:2007iy, Nishino:2009in, Farzinnia:2013pga, Guo:2015lxa, Kubo:2018kho, Burikham:2023bil}, inflation and late cosmology~\cite{Wu:2004rs, Ferrara:2010in, GarciaBellido:2011de, Kurkov:2013gma, Bars:2013yba, Csaki:2014bua, Kannike:2015apa, Kannike:2015kda, Salvio:2017xul, Ferreira:2018qss, Ghilencea:2018thl, Ghilencea:2019rqj, Ferreira:2019zzx, Ghilencea:2018dqd, Gunji:2019wtk, Ishiwata:2018dxg, Ishida:2019wkd, Barnaveli:2018dxo, Gialamas:2019nly, Karam:2018mft, Ghilencea:2020piz, Cai:2021png, Gialamas:2022xtt}. In these studies, discussions were mainly focused on two observables, spectral tilt $n_s-1$ for scalar perturbation and tensor-to-scalar ratio $r$ for primordial gravitational waves (GWs), which are constrained by the cosmic microwave background (CMB), $n_s=0.965\pm 0.009$ and $r<0.032$ at $95\%$ confidence level \cite{Planck:2018jri,Tristram:2021tvh}. 

Here we point out there are additional GWs at high frequency, which provide another probe of inflation dynamics with Weyl scaling symmetry~\cite{Wang:2022ojc, Wang:2023hsb}. Since the first direct detection of GWs \cite{LIGOScientific:2016aoc} and the recent observation of Hellings-Downs correlations which points to the gravitational-wave origin of this signal in the nano-Hz range \cite{NANOGrav:2023gor, NANOGrav:2023hvm, Antoniadis:2023rey, Antoniadis:2023zhi, Xu:2023wog, Reardon:2023gzh}, the era of multi-band GWs has arrived. GWs with different frequencies would be able to probe different physics. In the inflation scenario, a period called preheating could be present to bridge inflation and reheating before the Big Bang cosmology. Preheating can be realized by parameter resonance~\cite{Kofman:1994rk} in which the inflaton oscillates around the minimum of the potential, transfering energy to the coupled daughter fields. This period can cause significant energy transfer from inflaton to matter field, large and rapidly changing inhomogeneities, and GWs production \cite{Khlebnikov:1997di}. Besides, the self-resonance of inflaton around the minimum of potential can induce inhomogeneities of field configuration as well, which would also lead to the emission of GWs~\cite{Amin:2011hj,Zhou:2013tsa,Liu:2017hua}.
 
In the viable inflation models with Weyl scaling symmetry~\cite{Wang:2022ojc, Wang:2023hsb}, the second derivative of the nonconvex potential satisfies $V''<0$ near the minimum which can trigger the self-resonance of inflaton. It causes the perturbation modes in Fourier space with $k^2+V''<0$ experiencing instability, which leads to significant energy density inhomogeneities, and sources sizable GWs. Hence, GWs from self-resonance preheating can play an important role distinguishing from different models. In this paper, we explore preheating and GWs production in the inflation model~\cite{Wang:2023hsb} and study the evolution of inflaton field both analytically and numerically. Based on lattice simulation, we investigate the emission of stochastic background of GWs in detail.

The paper is organized as follows. In Section \ref{section:A}, we introduce the inflation model constructed from the Weyl scaling symmetry, including the general framework and inflation models. In section \ref{section:B}, we analyze the linear behavior of the perturbation and present the lattice simulation results of field evolution in cosmic expansion background. In section \ref{section:C}, we compute the energy spectrum of GWs from the preheating. Finally, we give our conclusion.

Throughout the paper, we set $\hbar=c=1$ in natural units, the reduced Planck mass $m_{pl}=\frac{1}{\sqrt{8\pi G}}\approx 2.435\times 10^{18}$ GeV.
We work in the Friedmann-Lemaitre-Robertson-Walker (FLRW) metric $ds^2=-dt^2+a^2\left(t\right)dx^idx^i$.

\section{Model}
\label{section:A}

We consider the following action~\cite{Tang:2020ovf}, describing a scalar field $\phi$ and a Weyl gauge field $W_\mu$,
\begin{equation}  \label{eq1}
  \mathcal{S}\equiv \int d^4x \mathcal{L}=\int d^4x \sqrt{-g}\left[\frac{m_{pl}^2}{2}F(\hat{R},\phi)-\frac{1}{2}\zeta D^{\mu}\phi D_{\mu}\phi-\frac{1}{4g^2_{W}}F_{\mu\nu}F^{\mu\nu}\right],
\end{equation}
where the covariant derivative on $\phi$ is defined by $D_{\mu}=\partial_{\mu}-W_{\mu}$, $g_W$ is the gauge coupling, and $F(\hat{R},\phi)$ is an arbitrary scaling invariant function of scalar $\phi$ and modified Ricci scalar $\hat{R}$ that defined by the local scaling invariant connection,
\begin{equation}\label{WGamma}
	\begin{aligned}	\hat{\Gamma}^\rho_{\mu\nu}&=\Gamma^\rho_{\mu\nu}+\left(W_\mu g^\rho_\nu+W_\nu g^\rho_\mu-W^\rho g_{\mu\nu}\right),
	\end{aligned}
\end{equation}
here $\Gamma^\rho_{\mu\nu}$ is the usual Christoffel connection, from which we have the relation between $\hat{R}$ and usual Ricci scalar $R$,
\begin{equation}
	\label{eqHR}
	\hat{R}=R-6W_{\mu}W^{\mu}-6\nabla_{\mu}W^{\mu}.
\end{equation}

The above action is invariant under the following local Weyl scaling transformation,
\begin{equation}\label{eq:scaling}
  \begin{aligned}
    &&g_{\mu\nu} &\rightarrow f^2\left(x\right)g_{\mu\nu}, \\
    &&W_{\mu} &\rightarrow W_{\mu}-\partial_{\mu}\text{ln}|f\left(x\right)|,\\
    &&\phi&\rightarrow f^{-1}\left(x\right)\phi,
  \end{aligned}
\end{equation}
where $f(x)$ is an arbitrary nonzero function. To simplify the formulation, we can introduce an auxiliary field $\chi$ and rewrite the Lagrangian as
\begin{equation}
  \mathcal{L}= \sqrt{-g}\left[\frac{m_{pl}^2}{2}F_{\hat{R}}(\chi^2,\phi) (\hat{R}-\chi^2)+\frac{1}{2}F(\chi^2,\phi)-\frac{1}{2}\zeta D^{\mu}\phi D_{\mu}\phi-\frac{1}{4g^2_{W}}F_{\mu\nu}F^{\mu\nu}\right],
\end{equation}
where $F_{\hat{R}}$ denotes the first derivative $\partial F/\partial\hat R$. The equivalence between two Lagrangian can be verified by applying the equation of motion for $\chi$. At this stage, Weyl scaling symmetry is still manifest under the transformation of Eq.~\ref{eq:scaling} and $\chi \rightarrow f^{-1}\left(x\right) \chi $. 

After fixing the gauge of scaling symmetry, 
\begin{equation}\label{eq:gaugefixing}
f=F_{\hat{R}}\left(\chi^2,\phi\right)=1,
\end{equation}
the resulting theory in Einstein frame describes an interacting massive vector field and a scalar field with the following potential
\begin{equation}
    V\left(\phi\right)=-\left[F(\chi^2,\phi)-\chi^2F_{\hat{R}}(\chi^2,\phi)\right]/2,
    \label{VF}
\end{equation}
where $\chi^2$ can be solved as a function of $\phi$ through above gauge fixing equation, Eq.~\ref{eq:gaugefixing}. 

The quadratic term $W_{\mu}W^{\mu}$ in Eq. \ref{eqHR} and the kinetic term of $\phi$ can be rewritten as
\begin{equation}
\label{eqgauge}
\frac{1}{2}\zeta D^{\mu}\phi D_{\mu}\phi+3W_{\mu}W^{\mu}=\frac{1}{2}\left(6+\zeta \phi^2\right)\overline{W}_{\mu}\overline{W}^{\mu}+\frac{1}{2}\frac{6\zeta}{6+\zeta\phi^2}\partial^{\mu}\phi\partial_{\mu}\phi,
\end{equation}
where $\overline{W}_{\mu}=W_{\mu}-\frac{1}{2}\partial_{\mu}\text{ln}\left|6+\zeta\phi^2\right|$ is a shift of field $W_{\mu}$ and would not change its kinetic term, $F_{\mu\nu}F^{\mu\nu}=\overline{F}_{\mu\nu}\overline{F}^{\mu\nu}$.
Besides, the first term in the right side of Eq. \ref{eqgauge} provides a mass term to
the vector field. And the second term is a non-canonical kinetic term which can be normalized as follows,
\begin{equation}
\frac{d\Phi}{d\phi} =\pm \sqrt{\frac{6\zeta}{6+\zeta\phi^2}} \Rightarrow 
 \begin{cases}
	\phi=\sqrt{\dfrac{6}{+\zeta}}\text{sinh}\dfrac{\pm\Phi}{\sqrt{6}}, & \zeta>0,\vspace{0.3cm} \\ 
	\phi=\sqrt{\dfrac{6}{-\zeta}}\text{cosh}\dfrac{\pm\Phi}{\sqrt{6}}, & \zeta<0.
\end{cases}
\end{equation}
Finally, we obtain the Lagrangian after normalization and redefinition,
\begin{equation}
    \frac{\mathcal{L}}{\sqrt{-g}}=\frac{m_{pl}^2}{2}R-\frac{1}{2}\partial^{\mu}\Phi\partial_{\mu}\Phi-V\left(\Phi\right)-\frac{1}{4g_{W}^2}\overline{F}_{\mu\nu}\overline{F}^{\mu\nu}-\frac{1}{2}m^2\left(\Phi\right)\overline{W}^{\mu}\overline{W}_{\mu} ,
\end{equation}
where $m^2\left(\Phi\right)=6+\zeta\phi^2$. The scalar $\Phi$ plays the role of inflaton and the massive gauge boson $\overline W_\mu$ is a dark matter candidate due to its $Z_2$ symmetry. The production mechanism of $\overline W_\mu$ has been discussed in~\cite{Wang:2022ojc}, hence we will not explore it further here. Since dark matter is sparse in the early universe, $\overline W_\mu$ hardly affects the evolution of the inflaton $\Phi$, we will only consider a single scalar field inflation model inspired by Weyl scaling invariant gravity in later discussions,
\begin{equation}
  \mathcal{S}=\int d^4x \sqrt{-g} \left[\frac{m_{pl}^2}{2}R-\frac{1}{2}\partial_{\mu}\Phi\partial^{\mu}\Phi-V\left(\Phi\right)\right].
\end{equation} 

\begin{figure}
	\centering
	\includegraphics[width=\textwidth]{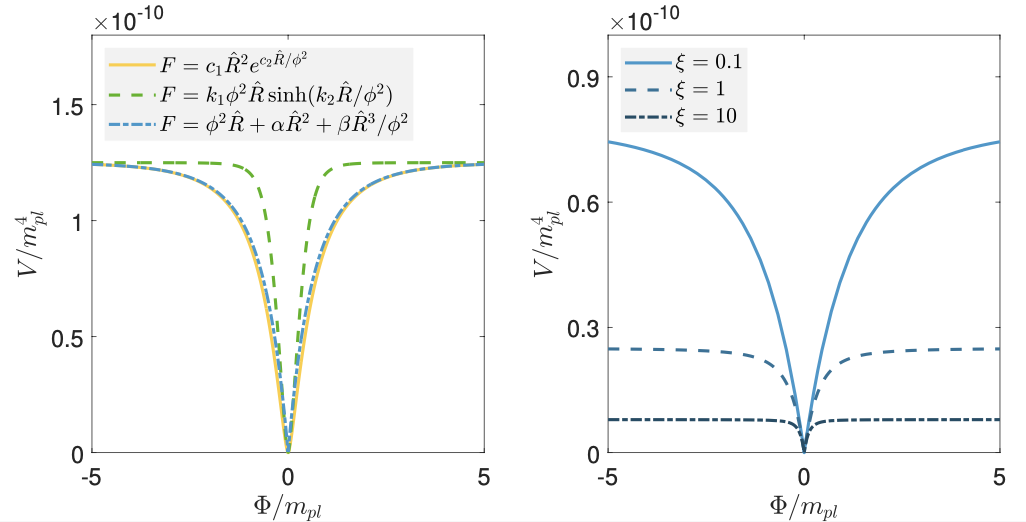}
	\caption{Left: several examples of nonconvex inflaton potential derived from Weyl $F(\hat R,\phi)$ gravity. The parameters are set as follows: $c_1=10^9$, $c_2=1$, $k_1=10^9$, $k_2=1$, $\alpha=10^9$, $\beta=10^9$, $\zeta=10^9$. Right: approximate potential in Eq.~\ref{eq5} with $\xi=0.1,~1,~10$, and $\alpha$ is chosen as $\alpha=5\times10^9\sqrt{\xi}$ obeying the CMB constraint. }
	\label{WeylFR}
\end{figure}

In~\cite{Wang:2023hsb} we have investigated the polynomial model with three terms,
\begin{equation}\label{eq:cubic}
F(\hat{R},\phi)=\phi^2\hat{R}+\alpha\hat{R}^2+{\beta}\hat{R}^3/{\phi^2},
\end{equation}
which gives viable inflation. Actually, there are many Weyl scaling invariant $F(\hat{R},\phi)$ models that can give rise to viable inflation, for instance,
\begin{equation}
F(\hat{R},\phi)=c_1 \hat R^2\exp({c_2\hat R}/{\phi^2}),\; k_1\phi^2\hat R\sinh({k_2\hat R}/{\phi^2}), \;\sum_{n>0}\alpha_n\phi^{4-2n}\hat R^n.
\end{equation} 
We illustrate their potential $V(\Phi)$ in the left panel of Fig.~\ref{WeylFR}. All these models have nonconvex potentials that can lead to the self-resonance of inflaton and production of GWs. Since the polynomial model Eq.~\ref{eq:cubic} can give an analytical expression of potential, without losing generality, we focus on it in this paper for discussion of GWs production.

The inflaton potential derived from Eq.~\ref{eq:cubic} is calculated to be
\begin{equation}
	\begin{aligned}
		V(\Phi)=\frac{\alpha}{6\beta}\left(\phi^4-\phi^2\right)&+\frac{\alpha^3\phi^4}{27\beta^2}\left[\left(1-\frac{3\beta}{\alpha^2}\left(1-\phi^{-2}\right)\right)^{3/2}-1\right],\\
        \mathrm{where}~&\phi^2=\frac{6}{\zeta}\sinh^2\left(\frac{\Phi}{\sqrt 6}\right),~\zeta>0.
		\label{VWR3}
	\end{aligned}
\end{equation}
When $\beta>0$, the potential has a minimum at $\Phi=0$, shown as the blue curve in the left part of Fig.~\ref{WeylFR}. The higher-order ($n\geq4$) terms of the polynomial model hardly affect the shape of the potential, except for making the potential more rounded at $\Phi=0$. So we mainly focus on the first three orders, and only consider the higher fourth order in the lattice simulation for numerical stability. When $\Phi$ is around its minimum, we can simplify Eq.~\ref{VWR3} as
\begin{equation}
  \label{eq5}
  \frac{V\left(\Phi\right)}{m_{pl}^4}\simeq -\frac{\xi\tilde{\Phi}^2}{2\alpha}+\frac{\xi^2\tilde{\Phi}^4}{3\alpha}\left[\left(1+\frac{1}{\xi\tilde{\Phi}^2}\right)^{3/2}-1\right],
\end{equation}
where $\tilde{\Phi}$ is a dimensionless field variable, $\tilde{\Phi}={\Phi}/{m_{pl}}$, and both $\alpha$ and $\xi={\alpha^2}/{(3\beta\zeta)}$ are dimensionless. We plot the potential in the right panel of Fig.~\ref{WeylFR}. This approximate potential can be used to study the preheating stage, in which the inflaton field $\Phi$ oscillates around the minimum of the potential at the end of inflation.

\begin{figure}[h]
	\centering
	\includegraphics[width=0.85\textwidth]{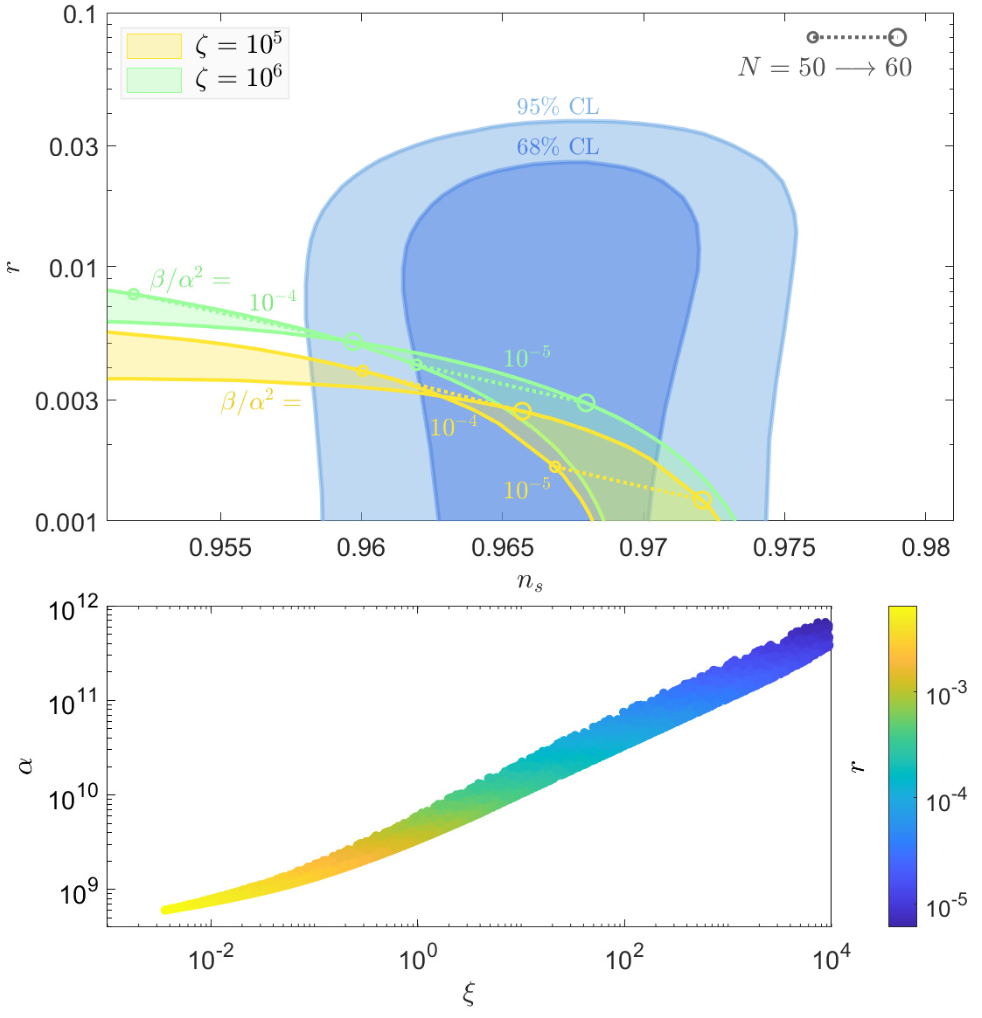}
	\caption{Observational constraints vs theoretical predictions in Weyl polynomial model Eq.~\ref{VWR3}. Upper: predictions of tensor-to-scalar ratio $r$ and spectral index $n_s$ with various parameters $\zeta$, $\beta$ and e-folding number $N\sim(50,~60)$. Blue area is the observational constraints given by the BICEP/Keck collaboration \cite{BICEP:2021xfz}. Lower: viable parameter space allowed by BICEP/Keck observations with $N=60$. Here we have redefined $\xi\equiv\frac{\alpha^2}{3\beta\zeta}$. The color gradient represents the tensor-to-scalar ratio $r$ obtained with the corresponding parameters.}
	\label{nsr}
\end{figure}

Now we discuss the observational constraints on the inflation model Eq.~\ref{VWR3}. There are two essential physical quantities for a specific model $V(\Phi)$, the tensor-to-scalar ratio $r=16\epsilon$, related to the magnitude of primordial gravitational wave, and the spectral index $n_s=1-6\epsilon+2\eta$ of primordial perturbations, where $\epsilon\equiv\frac{1}{2}\left[\frac{V'(\Phi)}{V}\right]^2$ and $\eta\equiv\frac{V''(\Phi)}{V}$ are slow-roll parameters. The upper panel of Fig.~\ref{nsr} shows some calculations of $r$ and $n_s$ with various parameters $\zeta$, $\beta$ and e-folding number $N\sim(50,~60)$ compared with the observation constraints given by the BICEP/Keck collaboration \cite{BICEP:2021xfz}. Here $\alpha$ is fitted by the observed value of the amplitude of scalar spectra $A_s\simeq\frac{V}{24\pi^2\epsilon}=2.1\times10^{-9}$. In the lower panel of Fig.~\ref{nsr}, we display the allowed parameter space of $N=60$ case with the parameter $\xi$. The color gradient represents $r$ obtained from the corresponding parameters. It is apparent that $\xi$ has a lower limit, $\xi\simeq4\times10^{-3}$, which corresponds to the upper limit on $r$ from observation. When $\xi>0.1$, the allowed $\alpha$ and $\xi$ approximately satisfy a simple relation, $\alpha\simeq5\times10^9\sqrt\xi$ (the coefficient is slightly smaller for $N=50$ case, about $3\times10^9$). Since $\alpha$ determines the height of the potential, $r$ is inversely proportional to $\sqrt\xi$ according to the relation $A_s=2.1\times10^{-9}$. Therefore, when $\xi$ is large enough, $r$ can be extremely small in this model.

\section{Dynamics}\label{section:B}

To describe the preheating dynamics after inflation, we list the equation of motion for scalar field $\Phi$ and Friedmann equations for scale factor $a$ in FLRW spacetime,
\begin{equation}
  \label{eq6}
  \ddot{\Phi}-\frac{1}{a^2}\nabla^2\Phi+3H\dot{\Phi}+\frac{\partial V}{\partial \Phi}=0,
\end{equation}
\begin{equation}
  \label{eq7}
  \frac{\ddot{a}}{a}=\frac{1}{3 m_{pl}^2}\langle -2 \frac{1}{2}\dot{\Phi}^2+V\left(\Phi\right)\rangle
\end{equation}
\begin{equation}
  \label{eq8}
  \left(\frac{\dot{a}}{a}\right)^2=\frac{1}{3m_{pl}^2}\langle \frac{1}{2}\dot{\Phi}^2+\frac{1}{2a^2}\sum_i \left(\partial_i\Phi\right)^2+V\left(\Phi\right)\rangle ,
\end{equation}
where the dot $\dot{}$ refers to derivative with respect to cosmic time $t$, $ \langle ... \rangle $ is average over space. In the following subsections, we will analyze the evolution of field configuration. By linear analysis, we can obtain the resonance spectrum of the corresponding $k$ mode. Then a detailed study of the dynamics is explored by lattice simulation.

\subsection{Linear Analysis}

To understand the nonlinear behavior of scalar field $\Phi$, we can do linear analysis for perturbation in the small inhomogeneous phase. We can expand field $\Phi\left(t,\bm{x}\right)$ as $\Phi\left(t,\bm{x}\right)=\Phi\left(t\right)+\delta\Phi\left(t,\bm{x}\right)$, where $\Phi\left(t\right)$ is homogeneous term, and $\delta\Phi\left(t,\bm{x}\right)$ is small perturbation.
Substituting the expansion into Eq. \ref{eq6}, keeping in first order perturbation, neglecting Hubble friction, we obtain
\begin{equation}
  \label{eq9}
  \ddot{\Phi}\left(t\right)+\frac{\partial V}{\partial \Phi}=0,
\end{equation}
\begin{equation}
  \label{eq10}
  \delta\ddot{\Phi}_{\bm{k}}+\left(k^2+\frac{\partial^2 V}{\partial^2\Phi}\right)\delta\Phi_{\bm{k}}=0
\end{equation}
where $\delta\Phi_{\bm{k}}$ is perturbation of momentum $\bm{k}$ in Fourier space.

Neglecting Hubble friction, the homogeneous term in Eq. \ref{eq9} describes the motion under conservative potential $V\left(\Phi\right)$. From Eq. \ref{eq10}, if $k^2+\frac{\partial^2 V}{\partial^2\Phi}>0$, $\delta\Phi_{\bm{k}}$ is in periodic behavior and the fluctuation do not increase significantly. However, if $k^2+\frac{\partial^2 V}{\partial^2\Phi}<0$ is satisfied over a long period of time, the fluctuations will increase exponentially, leading to inhomogeneous behavior of field configure. Then the perturbation phase is over, nonlinear phase is entered. The second derivative of potential from Weyl gravity in Eq. \ref{eq5} satisfies $\frac{\partial^2 V}{\partial^2\Phi}<0$. This means that $\bm{k}$ mode obeying $k^2+\frac{\partial^2 V}{\partial^2\Phi}<0$ will experience exponential growth. An illustration of potential $V\left(\Phi\right)$ is shown in Fig. \ref{WeylFR}.

\begin{figure}[t]
  \centering
  \begin{subfigure}[b]{0.48\textwidth}
    \includegraphics[width=\textwidth]{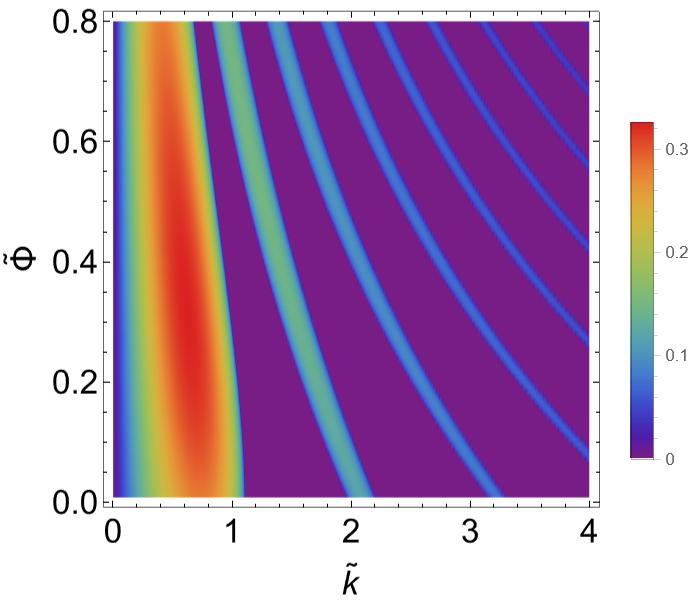}
    \caption{$\xi=1$}
    \label{fig:fig2subfig1}
  \end{subfigure}
  \hfill
  \begin{subfigure}[b]{0.48\textwidth}
    \includegraphics[width=\textwidth]{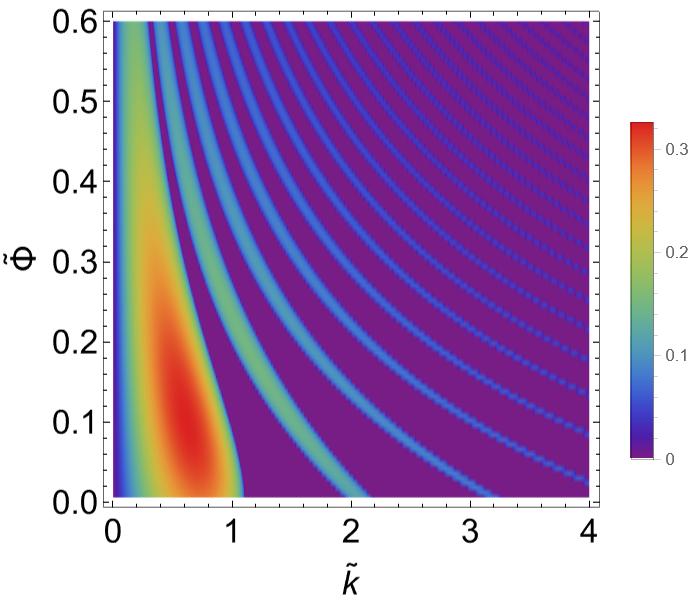}
    \caption{$\xi=10$}
    \label{fig:fig2subfig2}
  \end{subfigure}

  \caption{Resonance strength $Re\left(\mu_k\right)$ for the potential $V\left(\Phi\right)$ in 
  Eq. \ref{eq5} as a function of momentum $k$ and initial field value $\Phi_i$ with the colorbar 
  denoting the real part of the Floquet exponent $Re\left(\mu_k\right)$. $\alpha$ is 
  chosen the same as in Fig. \ref{WeylFR}. The corresponding momentum $k$ is normalizationg 
  by $\tilde{k}=\frac{k}{\omega}$, $\omega$ for different $\xi$ is given in Section \ref{section:lattice}. 
  The initial field value $\Phi_i$ is normalized by $\tilde{\Phi}_i=\frac{\Phi_i}{m_{pl}}$.} 
  \label{fig:fig2}
\end{figure}

According to Floquet theorem, the solution of Eq. \ref{eq10} can be written in the following form,
\begin{equation}
  \delta\Phi_{\bm{k}}\left(t\right)=\exp\left(\mu_{\bm{k}}\right)P_{+}\left(t\right)+\exp\left(-\mu_{\bm{k}}\right)P_{-}\left(t\right),
\end{equation}
where $\mu_{\bm{k}}$ is Floquet exponent and $P_{\pm}$ are periodic function. $Re\left(\mu_{\bm{k}}\right)  \neq 0$ can cause exponential growth of corresponding $\bm{k}$ mode.
Capturing resonance spectrum requires calculating $Re\left(\mu_{\bm{k}}\right)$ of different $\bm{k}$ modes and initial conditions. We plot the resonance strength as  a function of $k$
and $\Phi_i$ in Fig \ref{fig:fig2}.
There is a broad resonance region in each subfigure of Fig \ref{fig:fig2}, the perturbation of momentum $\bm{k}$ in this region will
experience a sharp enhancement. In an expanding backgroud, the physical momentum is $\frac{\bm{k}}{a}$. At early stage, small momentum modes $\tilde{k}\approx 1$ are in broad resonance region, and these modes will increase in a short time and then exit the resonance broad with the expansion of background. At later stage, 
as scale factor $a$ becomes larger, large $k$ modes will enter the resonance region, and be enhanced. With the decrease of field $\Phi$, the resonance region shrinks. A detailed study of full evolution process needs lattice simulation.

\subsection{Lattice Simulation}\label{section:lattice}

A number of codes have been developed to simulate the physics in the early Universe, such as LATTICEEASY\cite{Felder:2000hq}, CUDAEASY\cite{Sainio:2009hm}, PSpectRe\cite{Easther:2010qz}, CosmoLattice\cite{Figueroa:2020rrl,Figueroa:2021yhd}, etc. Since the inflaton potential derived from the the extensions with a small $\hat{R}^4$ correction with no explicit form, we develop our own codes to conduct the simulations using finite-difference method and the second-order symplectic algorithm.

In this section, we present the lattice simulation results on preheating in the inflation model Eq.~\ref{eq:cubic}. We introduce some variables to make physical quantities dimensionless, the field configuration $\Phi$ is divided by $m_{pl}$, $\tilde{\Phi}=\frac{\Phi}{m_{pl}}$ and
\begin{equation}
  d\tilde{\eta}= a^{-1} \omega dt,\;
  \tilde{k}=\frac{k}{\omega},
\end{equation}
where $\omega$ is chosen as the frequency of the oscillation of the homogeneous mode of inflaton field $\Phi$, and time coordinate is chosen to be $\tilde{\eta}$ for two considerations. The factor $a^{-1}$ can make the average of the field configure a stable period. $\omega$ multiplied in the expression can make the time coordinate dimensionless and the evolution period an $\mathcal{O}\left(1\right)$ quantity.
We need to choose different $\omega$ for different $\xi$.
\begin{eqnarray}
&&\omega=1.82\times10^{13}\; \text{GeV} \quad \text{for} \; \xi=1,   \nonumber  \\
&&\omega=2.19\times10^{13}\; \text{GeV} \quad \text{for} \; \xi=10.
\end{eqnarray}
The lattice size is chosen to be $N=256$, and the infrared momentum $k_{\text{IR}}$ depends on the peak of field configuration in momentum space.

Initial condition should be set for the simulation. Initial field configure $\Phi$ and its first order derivative are set by a homogeneous term and fluctuations. The homogeneous term $\tilde{\Phi}_{i}$ and its derivative is set at the end of inflation,
\begin{eqnarray}
&&\tilde{\Phi}_{i}=0.75 \quad \text{and} \quad \tilde{\Phi}_{i}^{\prime}=8.7\times10^{-7} \quad \text{for}\; \xi=1,   \nonumber\\
&&\tilde{\Phi}_{i}=0.51 \quad \text{and} \quad \tilde{\Phi}_{i}^{\prime}=2.5\times10^{-7} \quad \text{for}\; \xi=10.
\end{eqnarray}
where $\prime$ denotes derivative with respect to $\tilde{\eta}$.
The initial fluctuations are obtained from quantum vacuum fluctuation. We set the initial scale factor $a_i=1$, and
its derivative is determined by Friedmann equation. For a detailed description of the numerical method, see Appendix \ref{appen:A}.

We solve Eq. \ref{eq6} and Eq. \ref{eq7} numerically with lattice simulation and treat Eq.~\ref{eq8} as a constraint which should be satisfied in the whole evolution. In our simulation, we introduce a ${\delta}\hat{R}^4/{\phi^4}$ term in the Weyl gravity with a small parameter ${\delta}$, which can provide a smooth transition at the minimum and help to avoid the singularity behavior in potential $V\left(\Phi\right)$ in Eq. \ref{eq5}.

\begin{figure}[t]
  \centering

    \includegraphics[width=0.48\textwidth]{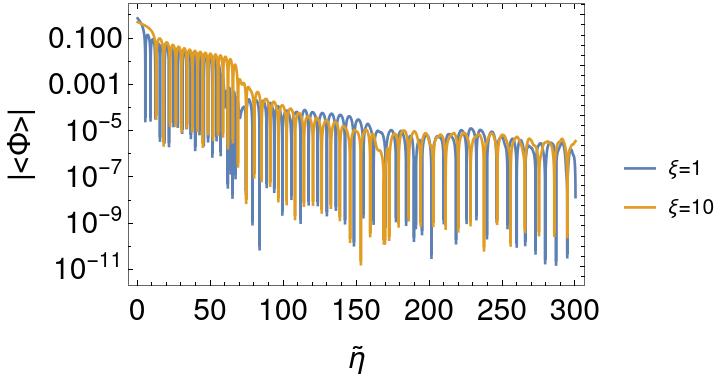}
    \includegraphics[width=0.48\textwidth]{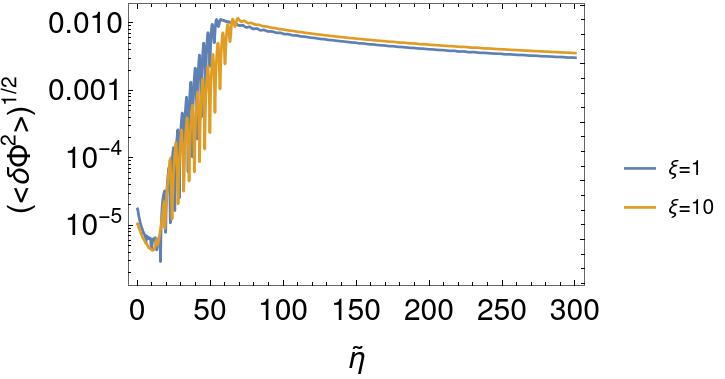}
    \includegraphics[width=0.48\textwidth]{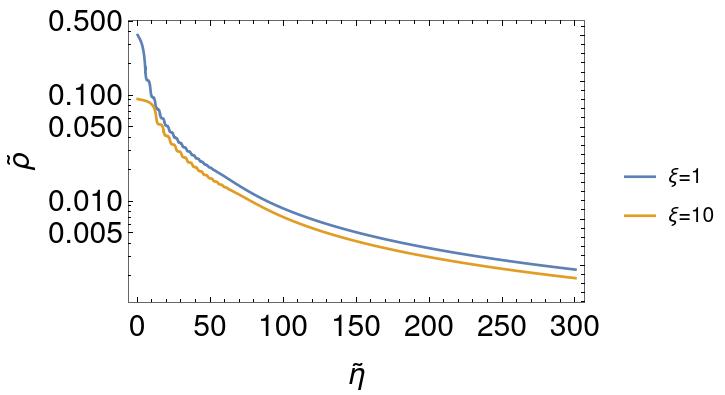}
    \includegraphics[width=0.48\textwidth]{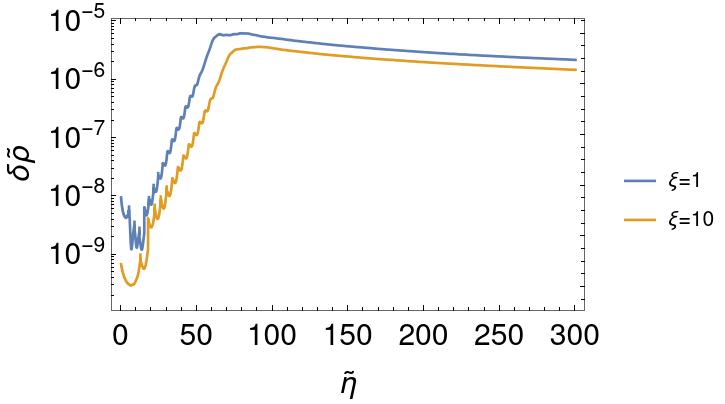}

  \caption{Evolution of absolute value of the average field value $\Phi$, the root-mean-squared of field fluctuation $\left(<\delta\Phi^2>\right)^{1/2}$, the energy density $\tilde{\rho}$ and the energy density of fluctuation $\left(<\delta\tilde{\rho}^2>\right)^{1/2}$.
   The energy is $\rho$ is divided by $\omega^2m_{pl}^2$, $\tilde{\rho}=\frac{\rho}{\omega^2m_{pl}^2}$. The blue line captures the case when 
   $\xi=1$ and the yellow line captures the case when $\xi=10$.} 
  \label{fig:fig3}
\end{figure}

Fig.~\ref{fig:fig3} shows the evolution of the averaged field value, the root-mean-squared of field fluctuation, the energy density and the energy density of fluctuation. The mean of the scalar field decreases because of the Hubble friction. A short time after the beginning, around $\tilde{\eta}=20  \sim 100$, the field fluctuation grows exponentially since the resonance of corresonding $k$ modes. The growth of fluctuation stops at $\tilde{\eta}\approx 100$  until the back-reaction effects become remarkable. At that time, the linear approximation breaks down and the evolution enters the nonlinear stage. Fig. \ref{fig:fig3} shows sizable field fluctuation and energy fluctuation growth, which can cause significant GWs signals.

\section{Gravitational Waves Production}\label{section:C}

In this section, we study GWs production from the corresponding preheating after inflation. Consider the  metric with
 the tensor perturbation,
 \begin{equation}
  \label{eq14}
  ds^2=-dt^2+a^2\left(\delta_{ij}+h_{ij}\right)dx^i dx^j,
 \end{equation}
where $h_{ij}$ satisfies transverse-traceless condition, $\partial_ih_{ij}=0$ and $\Sigma_i h_{ii}=0$. Substituting Eq. \ref{eq14} in Einstein equation and keeping in 
first order of $h_{ij}$, we obtain
\begin{equation}
  \label{eq15}
  \ddot{h}_{ij}+3H\dot{h}_{ij}-\frac{1}{a^2}\nabla^2 h_{ij}=\frac{2}{m^2_{pl}}\Pi_{ij}^{TT},
\end{equation}
where $\Pi_{ij}^{TT}$ is the transverse-traceless part of the anisotropic stress tensor $\Pi_{ij}=\partial_i\Phi\partial_j\Phi$.

In our implementation of lattice simulation, we do not solve Eq. \ref{eq15} directly. Instead, we solve the following equation \cite{Bethke:2013vca},
\begin{equation}
  \ddot{u}_{ij}+3H\dot{u}_{ij}-\frac{1}{a^2}\nabla^2 u_{ij}=\frac{2}{m^2_{pl}}\Pi_{ij}.
\end{equation}
Only when we need to compute gravitational spectrum, we can extract tensor perturbation $h_{ij}$ from $u_{ij}$ by the projection operator,
\begin{equation}
  \Lambda_{ij,lm}\left(\hat{\bm{k}}\right)=P_{il}\left(\hat{\bm{k}}\right)P_{jm}\left(\hat{\bm{k}}\right)-\frac{1}{2}P_{ij}\left(\hat{\bm{k}}\right)P_{lm}\left(\hat{\bm{k}}\right),
\end{equation}
where $P_{ij}=\delta_{ij}-\hat{\bm{k}}_i\hat{\bm{k}}_j$, and $\hat{\bm{k}}_i=\frac{\bm{k}_i}{k}$. In momentum space, the tensor perturbation can be derived by $h_{ij}\left(\bm{k},t\right)=\Lambda_{ijkl}\left(\bm{k}\right)u_{kl}\left(\bm{k},t\right)$.

\begin{figure}[t]
  \centering
  \begin{subfigure}[b]{0.48\textwidth}
    \includegraphics[width=\textwidth]{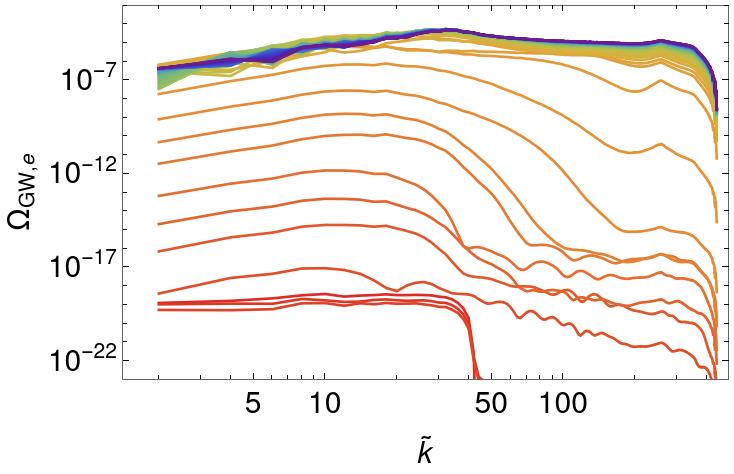}
    \caption{$\xi=1$}
    \label{fig:fig5subfig1}
  \end{subfigure}
  \hfill
  \begin{subfigure}[b]{0.48\textwidth}
    \includegraphics[width=\textwidth]{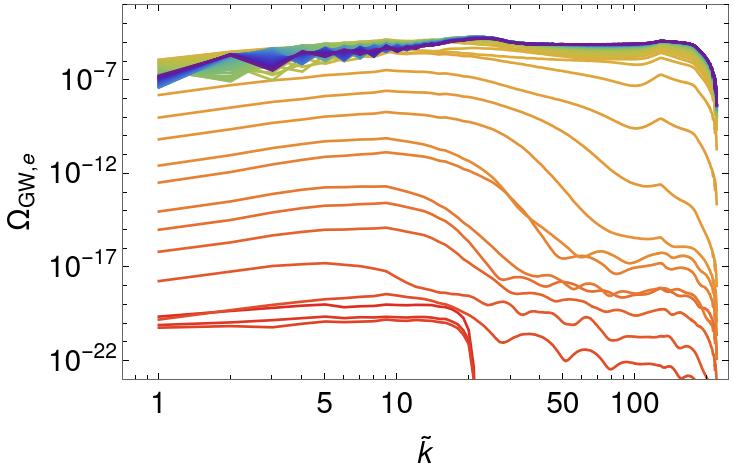}
    \caption{$\xi=10$}
    \label{fig:fig5subfig2}
  \end{subfigure}

  \vspace{0.5cm}

  \begin{subfigure}[b]{0.48\textwidth}
    \includegraphics[width=\textwidth]{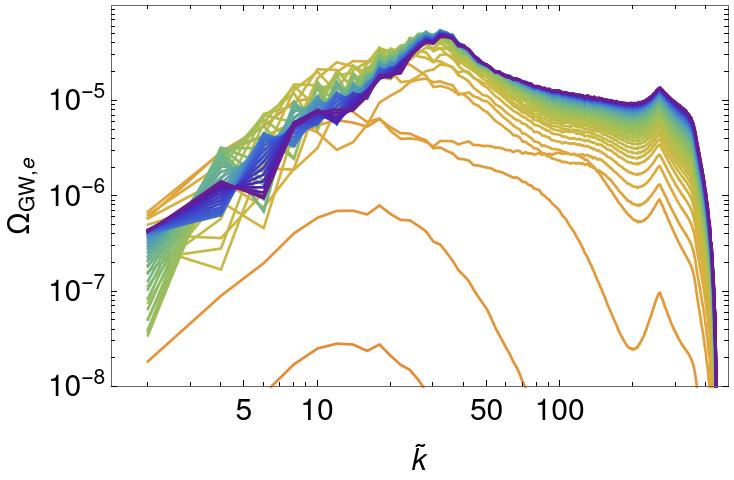}
    \caption{$\xi=1$}
    \label{fig:fig5subfig3}
  \end{subfigure}
  \hfill
  \begin{subfigure}[b]{0.48\textwidth}
    \includegraphics[width=\textwidth]{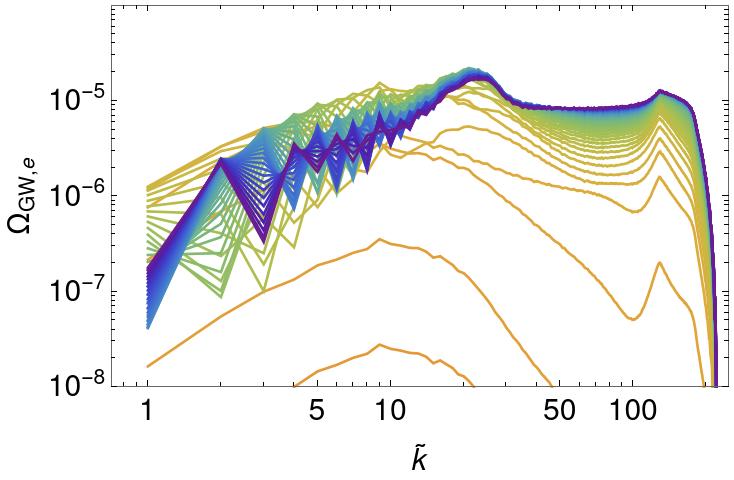}
    \caption{$\xi=10$}
    \label{fig:fig5subfig4}
  \end{subfigure}
  \caption{The evolution of GWs energy spectrum with each curve as a snapshot. The top panels show the time evolution 
  of GWs spectra when $\xi=1$ and $\xi=10$. The spectra are plotted at time $\tilde{\eta}=0, 5, 10,...$, and go from red (early-times) to blue (late-times).
  The bottom figures are zoomed to observe the peak better.} 
  \label{fig:fig5}
\end{figure}

The energy density of stochastic GW background is 
\begin{equation}
  \rho_{\text{GW}}=\frac{m_{pl}^2}{4}<\dot{h}_{ij}\dot{h}_{ij}>,
\end{equation}
where $<...>$ denotes spatial average. The energy density power spectrum of GWs is given by
\begin{equation}
  \Omega_{\text{GW}}\left(k,t\right)=\frac{1}{\rho_c}\frac{d\rho_{\text{GW}}}{d \text{log}k}\left(k,t\right),
\end{equation}
where $\rho_c=3m_{pl}^2H^2$ is the critical energy density of the Universe. In Fig.~\ref{fig:fig5}, we show the evolution of GWs energy spectrum $\Omega_{\text{GW}}$ in our simulation, where each curve corresponds to a snapshot at some time.

\begin{figure}[t]
  \centering
  \begin{subfigure}[b]{0.49\textwidth}
    \includegraphics[width=\textwidth,height=0.8\textwidth]{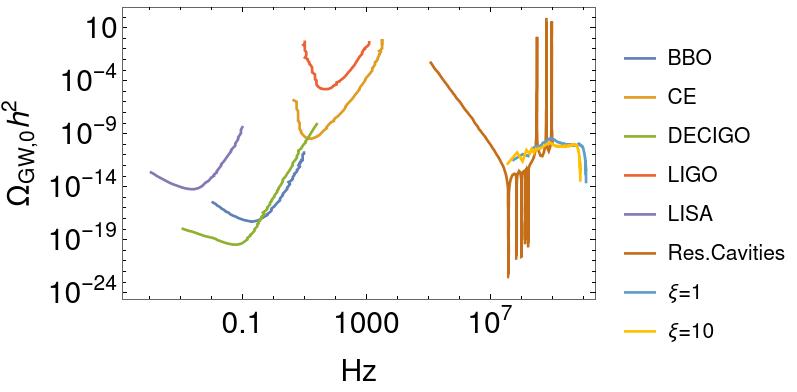}
    \label{fig:fig6subfig1}
  \end{subfigure}
  \hfill
  \begin{subfigure}[b]{0.49\textwidth}
    \includegraphics[width=\textwidth,,height=0.8\textwidth]{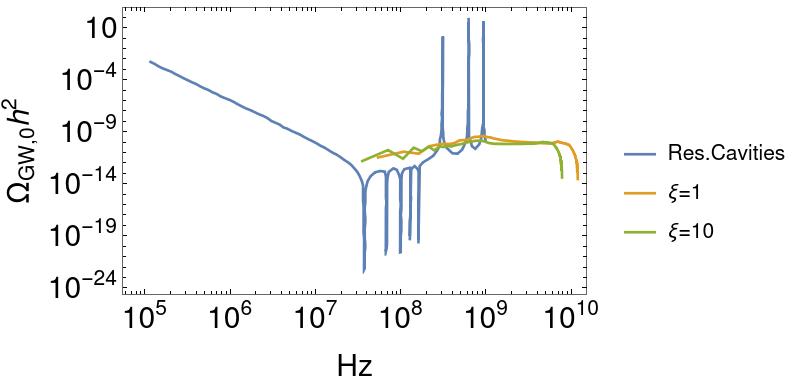}
    \label{fig:fig6subfig2}
  \end{subfigure}
  \caption{GWs energy spectrum today for preheating in Weyl gravity for $\xi=1$ and $\xi=10$. The sensitivity of resonance cavities is shown as the brown curve\cite{Herman:2020wao, Herman:2022fau}. For comparison, we also show the sensitivities of other GW experiments in lower frequencies, LISA and BBO~\cite{Thrane:2013oya}, DECIGO~\cite{Seto:2001qf}, LIGO~\cite{KAGRA:2013rdx} and CE~\cite{Reitze:2019iox}.} 
  \label{fig:fig6}
\end{figure}

Eventually, we need to obtain the GW energy spectra today. We can calculate as \cite{Dufaux:2007pt},
\begin{equation}
  f=\frac{k_0}{2\pi}=\frac{k}{a_e\rho_e^{1/4}}\left(\frac{a_e}{a_*}\right)^{1-\frac{3}{4}\left(1+\omega_*\right)} 4\times10^{10}\text{Hz},
\end{equation}
\begin{equation}
  \Omega_{\text{GW,0}}=\Omega_{r,0}\left(\frac{g_0}{g_*}\right)^{1/3}\left(\frac{a_e}{a_*}\right)^{1-3\omega_*}\Omega_{\text{GW,e}},
\end{equation}
where $a_e$ denotes the scale factor $a$ at the end of simulation, $\omega_*$ is the equation of state parameter between $t_e$ and $t_*$, $g$ is the number of effective massless degrees of freedom (We have neglected the difference between $g$ and $g_S$.) we will take $g_*/g_0=100$ and $a_e=a_*$.
In Fig. \ref{fig:fig6}, we present the GWs energy density spectrum today. From Fig. \ref{fig:fig6}, we observe that the GWs signals peak at around $10^8$ Hz $\sim$ $10^9$ Hz. A higher frequency peak at $10^9$ Hz $\sim$ $10^{10}$ Hz comes from numerical noises, which is not a physical signal.

Searches for such high-frequency GWs are actively discussed~\cite{Liu:2023mll,Ito:2023fcr,Ito:2023nkq} recently, although probing GWs from preheating would require significant improvement on sensitivities. There are various proposals aimed at high-frequency GWs~\cite{Aggarwal:2020olq}, such as optically levitated sensors \cite{Aggarwal:2020umq}, detectors based on the inverse-Gertsenshtein effect~\cite{1962Wave, Braginskii:1973vm}, mechanical resonators~\cite{Goryachev:2014nna,Goryachev:2014yra,Aguiar:2010kn,Gottardi:2007zn}, microwave and optical cavities~\cite{Mensky:2009zz,Caves:1979kq}, interferometers~\cite{Ackley:2020atn, Bailes:2019oma, Akutsu:2008qv}, superconducting rings~\cite{CMB-S4:2020lpa} and the magnon modes~\cite{Ito:2019wcb, Ito:2022rxn}. It is noticeable that GWs from preheating in Weyl gravity can be tested by the proposed GWs dectector using resonance cavities~\cite{Herman:2020wao, Herman:2022fau}.

\section{Conclusion}

Cosmic inflation in the early universe provides a natural explanation to the horizon problem and flatness problem, and physical origin of the primordial perturbation. Motivated by the near scaling invariance during inflation, we have investigated the inflationary dynamics in inflation models with Weyl scaling symmetry, which are in good agreements with current observations and provide a dark matter candidate. Due to the nonconvex scalar potentials, such inflation models would generally lead to preheating that can cause large inhomogenerous and source strong GWs. We have analyzed the instability charts for different viable parameters and conducted numerical lattice simulations for the nonlinear dynamics. In the end, we have also computed the produced GWs at the present time. Sizable GWs around $10^8$ Hz $\sim$ $10^9$ Hz are produced due to the inhomogeneities, which may be observed by future high-frequency GWs detectors such as resonance cavities. Therefore, high-frequency GWs can provide an additional test for inflation models based on Weyl symmetry.

\section*{Acknowledgements}
\noindent Y.T. is supported by the National Key Research and Development Program of China under Grant No.2021YFC2201901, the National Natural Science Foundation of China (NSFC) under Grant No.~12147103 and the Fundamental Research Funds for the Central Universities. The work of W.Y.H and Y.Q.M. was supported by the National Natural Science Foundation of
China (Grants No. 11975029, No.12325503), the National Key Research and Development Program of China under
Contracts No. 2020YFA0406400, and the High-performance Computing Platform of Peking University. 

\appendix

\section{Numerical Method}
\label{appen:A}
In this section, we summarize the numerical details of our simulation.

\subsection{Initial Condition}

Here we show the initial condition in our numerical simulation. The initial field configuration $\Phi\left(t_i,\bm{x}\right)$ and momentum $\Phi^{\prime}\left(t_i,\bm{x}\right)$ are set by a homogeneous term and perturbation components. The homogeneous term is given at the violation of slow-roll condition, $\epsilon=1$ or $\eta=1$. We obtain the perturbation components following \cite{Figueroa:2020rrl}, 
\begin{equation}
\langle\delta\Phi^2\rangle=\int d \;\text{log}\; k \; \Delta_{\delta\Phi}\left(k\right), 
\Delta_{\delta\Phi}\left(k\right)=\frac{k^3}{2\pi^2}\mathcal{P}_{\delta\Phi}\left(k\right), \langle\delta\Phi_{\vec{k}}\delta\Phi_{\vec{k^{\prime}}}\rangle=\left(2\pi\right)^3\mathcal{P}_{\delta\Phi}\left(k\right)\delta(\vec{k}-\vec{k^{\prime}}),
\end{equation}
where $\langle ...\rangle$ represents an ensemble average, $\Delta_{\delta\Phi}\left(k\right)$ is the power spectrum. The initial condition is set by quantum vacuum fluctuations,
\begin{equation}
\label{eq:powerspectrum}
    \mathcal{P}_{\delta\Phi}\left(k\right)=\frac{1}{2a^2\omega_{k,\Phi}}, \quad
    \omega_{k,\Phi}=\sqrt{k^2+a^2 m^2_{\Phi}},\quad
    m^2_{\Phi}=\frac{\partial^2 V}{\partial^2\Phi}|_{\Phi=\overline{\Phi}},
\end{equation}
where $\omega_{k,\Phi}$ is the frequency of the $k$ mode, $m_{\Phi}$ is the effective mass of the scalar field.

In order to realize such an initial conditon on lattice, we consider a cubic lattice with lengh $L$ and sites $N$ per dimension. The gridpoints are labeled as
\begin{equation}
    \bm{n}=\left(n_1,n_2,n_3\right), n_i=0,1,...,N-1, \text{and } i=1,2,3.
\end{equation}
The space interval between adjacent gridpoints is $\Delta x=L/N$. The reciprocal lattice is given by
\begin{equation}
    \tilde{\bm{n}}=\left(\tilde{n}_1,\tilde{n}_2,\tilde{n}_3\right), \tilde{n}_i=-\frac{N}{2}+1,-\frac{N}{2}+2,...,-1,0,1,...,\frac{N}{2}-1,\frac{N}{2}, \text{and } i=1,2,3.
\end{equation}
Then we can define the discrete Fourier transformation as
\begin{equation}
    f\left(\bm{n}\right)=\frac{1}{N^3}\sum_{\tilde{\bm{n}}}e^{-i\frac{2\pi}{N}\tilde{\bm{n}}\bm{n}}f\left(\tilde{\bm{n}}\right).
\end{equation}
To mimic the spectrum in Eq. \ref{eq:powerspectrum}, we set the fluctuations on each grid point as
\begin{equation}
    \delta\Phi\left(\tilde{\bm{n}}\right)=\frac{1}{\sqrt{2}}\left(\left|\delta\Phi^{\left(l\right)}\left(\tilde{\bm{n}}\right)\right|e^{i\theta^{\left(l\right)}\left(\tilde{\bm{n}}\right)}+\left|\delta\Phi^{\left(r\right)}\left(\tilde{\bm{n}}\right)\right|e^{i\theta^{\left(r\right)}\left(\tilde{\bm{n}}\right)}\right),
\end{equation}
\begin{equation}
    \delta\Phi^{\prime}\left(\tilde{\bm{n}}\right)=\frac{1}{a^2}\left[\frac{i\omega_k}{\sqrt{2}}\left(\left|\delta\Phi^{\left(l\right)}\left(\tilde{\bm{n}}\right)\right|e^{i\theta^{\left(l\right)}\left(\tilde{\bm{n}}\right)}-\left|\delta\Phi^{\left(r\right)}\left(\tilde{\bm{n}}\right)\right|e^{i\theta^{\left(r\right)}\left(\tilde{\bm{n}}\right)}\right)\right]-\mathcal{H}\delta\Phi\left(\tilde{\bm{n}}\right),
\end{equation}
where we have used $d\eta=a\,dt$ and $\mathcal{H}=\frac{a^{\prime}}{a}$. In these expressions, $\theta^{\left(l\right)\left(\tilde{\bm{n}}\right)}$ and $\theta^{\left(r\right)\left(\tilde{\bm{n}}\right)}$ are two random phases varying uniformly in the range $\left[0,2\pi\right)$, $\delta\Phi^{\left(l\right)}\left(\tilde{\bm{n}}\right)$ and $\delta\Phi^{\left(r\right)}\left(\tilde{\bm{n}}\right)$ are random variables according to Rayleigh distribution,
\begin{equation}
    \left|\delta\Phi\left(\tilde{\bm{n}}\right)\right|^2=\left(\frac{N}{\Delta x}\right)^3\frac{1}{2a^2\sqrt{k^2\left(\tilde{\bm{n}}\right)+a^2m_{\Phi}^2}}.
\end{equation}
Besides, We set the initial scale factor $a_i=1$, and the first derivative of $a$ is constrained by the Friedmann equation.

\subsection{Integration Scheme}

We use the second order velocity-Verlet algorithm to solve the dynamics of preheating. Firstly, we use
parameters $\omega$ and $f_{*}=m_{pl}$ to make physical variables dimensionless,
\begin{equation}
  \tilde{\Phi}=\frac{\Phi}{f_{*}},\quad \tilde{\Phi^{\prime}}=\frac{1}{f_{*}\omega}\Phi^{\prime},
  \quad \Delta\tilde{x}_{\mu}=\omega \Delta x_{\mu}, \quad \tilde{V}\left(\tilde{\Phi}\right)=\frac{1}{\omega^2f_{*}^2}V\left(\Phi\right),
\end{equation}
where $x_0 =\eta$ and $d\eta=a\, dt$.
We define the forward and backward derivatives as,
\begin{equation}
  \Delta_{\mu}^{\pm}f\left(x\pm\Delta x_{\mu}/2\right)=\frac{\pm f\left(x \pm \Delta x_{\mu}\right)\mp f\left(x\right)}{\Delta x_{\mu}}.
\end{equation}
The dynamical variables are field configuration $\tilde{\Phi}$, momentum $\Pi=a^4\tilde{\Phi^{\prime}}$, scale factor $a$ and its first derivative $b=\frac{1}{\omega}\frac{da}{d\eta}$.

We evolve the momentum $\Pi$ firstly,
\begin{equation}
  \Pi\left(\tilde{\eta}+\frac{\Delta\tilde{\eta}}{2}\right)=\Pi\left(\tilde{\eta}\right)+\left(\Sigma_k \tilde{\Delta}^{-}_{k}\tilde{\Delta}^{+}_k\tilde{\Phi}\left(\tilde{\eta}\right)-a^2\left(\tilde{\eta}\right)\tilde{V}_{,\tilde{\Phi}}\left(\tilde{\eta}\right)\right)\frac{\Delta\tilde{\eta}}{2},
\end{equation}
Then we evolve the factor $b$,
\begin{equation}
  b\left(\tilde{\eta}+\frac{\Delta\tilde{\eta}}{2}\right)=b\left(\tilde{\eta}\right)+\frac{1}{3}\left(\frac{f_*}{m_{pl}}\right)^2 a^{-1}\left(\tilde{\eta}\right)\left(-3\tilde{E}_K-\tilde{E}_G\right)_{\tilde{\eta}}\frac{\tilde{\eta}}{2},
\end{equation}
where the average of kinetic energy $\tilde{E}_K=\langle\Pi^2\rangle/(2a^6)$, and the average of gradient energy $\tilde{E}_G=\Sigma_k\langle(\tilde{\nabla_k\tilde{\Phi}})^2\rangle/(2a^2)$.
Then we can obtain scale factor $a\left(\tilde{\eta}+\Delta\tilde{\eta}\right)$,
\begin{equation}
  a\left(\tilde{\eta}+\Delta\tilde{\eta}\right)=a\left(\tilde{\eta}\right)+b\left(\tilde{\eta}+\frac{\Delta\eta}{2}\right)\Delta\tilde{\eta},
\end{equation}
and
\begin{equation}
  a\left(\tilde{\eta}+\frac{\Delta\tilde{\eta}}{2}\right)=\frac{a\left(\tilde{\eta}\right)+a\left(\tilde{\eta}+\Delta\tilde{\eta}\right)}{2}.
\end{equation}
The field configuration at the next time slice can be obtained by
\begin{equation}
  \tilde{\Phi}\left(\tilde{\eta}+\Delta\tilde{\eta}\right)=\tilde{\Phi}\left(\tilde{\eta}\right)+a^{-4}\left(\tilde{\eta}+\frac{\Delta\tilde{\eta}}{2}\right)\Pi\left(\tilde{\eta}+\frac{\Delta\tilde{\eta}}{2}\right)\Delta\tilde{\eta}.
\end{equation}
We can evolve the momentum at the time $\tilde{\eta}+\Delta\tilde{\eta}$,
\begin{align}
&\Pi\left(\tilde{\eta}+\Delta\tilde{\eta}\right)=\Pi\left(\tilde{\eta}+\frac{\Delta\tilde{\eta}}{2}\right)+\left(\Sigma_k \tilde{\Delta}^{-}_{k}\tilde{\Delta}^{+}_k\tilde{\Phi}\left(\tilde{\eta}+\Delta\tilde{\eta}\right)-a^2\left(\tilde{\eta}+\Delta\tilde{\eta}\right)\tilde{V}_{,\tilde{\Phi}}\left(\tilde{\eta}+\Delta\tilde{\eta}\right)\right)\frac{\Delta\tilde{\eta}}{2},\nonumber \\
&b\left(\tilde{\eta}+\Delta\tilde{\eta}\right) = b\left(\tilde{\eta}+\frac{\Delta\tilde{\eta}}{2}\right)+\frac{1}{3}\left(\frac{f_*}{m_{pl}}\right)^2 a^{-1}\left(\tilde{\eta}+\Delta\tilde{\eta}\right)\left(-3\tilde{E}_K-\tilde{E}_G\right)_{\tilde{\eta}+\Delta\tilde{\eta}}\frac{\Delta\tilde{\eta}}{2}.
\end{align}

As mentioned in Section \ref{section:lattice}, We introduce a $\hat{R}^4$ term with a small parameter $\delta$ to avoid the singularity at the bottom of the scalar potential,
\begin{equation}
  F\left(\hat{R},\phi\right)=\phi^2\hat{R}+\alpha\hat{R}^2+\frac{\beta}{\phi^2}\hat{R}^3+\frac{\delta}{\phi^4}\hat{R}^4,
\end{equation}
We can derive the scalar potential
\begin{equation}
  V\left(\phi\right)=\frac{\alpha}{2}\chi^2+\frac{\beta}{\phi^2}\chi^6+\frac{3\delta}{2\phi^4}\chi^8,
\end{equation}
where the gauge fixing condition tells us the relation between $\phi$ and $\chi$
\begin{equation}
  \label{eq:relation}
  \phi^2+2\alpha\chi^2+3\beta\chi^4/\phi^2+4\delta\chi^6/\phi^4=1,
\end{equation}
the concrete form of the potential is tedious, A better way to compulate the potential or its derivative is as following. 
Since a small $\delta$ correction, we can use the approximate relation between $\phi$ and $\Phi$ in Ref. \cite{Wang:2023hsb},
\begin{equation}
  \phi^2=\frac{\Phi^2}{\zeta}.
\end{equation}
Then we can express the derivative of $V\left(\Phi\right)$ as 
\begin{equation}
  \frac{dV}{d\Phi}=\frac{d\phi}{d\Phi}\left(\frac{\partial V}{\partial \chi}\frac{d \chi}{d \phi}+\frac{\partial V}{\partial \phi}\right),
\end{equation}
where $\frac{d\chi}{d\phi}$ can be calculated from Eq. \ref{eq:relation}.

\bibliographystyle{unsrt}
\bibliography{ref}

\end{document}